\documentclass[epsfig,aps,fleqn,amsmath]{article}

\newcommand {\be}{\begin{equation}}
\newcommand {\ee}{\end{equation}}
\newcommand {\bea}{\begin{eqnarray}}
\newcommand {\eea}{\end{eqnarray}}
\newcommand {\refeq}[1] {(\ref{#1})}

\newcommand {\vett}[1] {\mathbf{#1}} 

\begin{document}


\title{Collective excitations of a trapped degenerate Fermi gas}

\author{ M. Amoruso, I. Meccoli\thanks{Permanent address: Dipartimento di Fisica, Universit\`a di Parma, Parco Area delle Scienze 7a, I-43100 Parma, Italy},
  	A. Minguzzi, M. P. Tosi\\
	Scuola Normale Superiore and \\
	Istituto Nazionale per 
	la Fisica della Materia,I-56126 Pisa,  Italy}

\date{\today}
\maketitle

\begin{abstract}
We evaluate the small-amplitude excitations of a spin-polarized vapour
of  Fermi atoms confined inside a harmonic trap. The dispersion law 
$\omega=\omega_{f}[l+4n(n+l+2)/3]^{1/2}$ is obtained for the vapour 
 in the collisional regime inside a spherical trap of frequency
$\omega_{f}$, with $n$ the number of radial nodes and $l$ the orbital angular
momentum. The low-energy excitations are also treated in the case of
an  axially  symmetric harmonic confinement. The collisionless regime
is discussed with main reference to a Landau-Boltzmann equation for
the  Wigner distribution function: this equation is solved within a 
 variational approach allowing an account for non-linearities. A
comparative  discussion of the eigenmodes of oscillation for confined
Fermi and Bose vapours is presented in an Appendix.
\end{abstract}

Pacs:67.40 Db Quantum statistical theory; ground state, elementary 
excitations.

\section{Introduction}

The achievement of condensation in confined vapours of bosonic alkali atoms \cite{anderson}- \cite{bradley} and more recently in vapours of Hydrogen \cite{fried} has been giving great stimulus to the study of many-body and quantum statistical effects in dilute quantal fluids. In particular, the observation of small-amplitude shape-deformation modes in these bosonic condensates at very low temperature \cite{jin}-\cite{mewes} has provided a crucial test for the theory  as developed by many authors (see for istance \cite{stringari_hydro_1996}-\cite{perez-garcia}). Trapping of fermionic  species has also been reported for 
$^{6}$Li \cite{mc-alex} and $^{40}$K \cite{cataliotti}. Cooling of these Fermi gases into the regime of quantal degeneracy still remains to be  achieved experimentally, but one may already anticipate that the observation of the dynamic response of the vapour of fermionic atoms to modulation of the trap will provide an important method of diagnostics. It is to the theoretical study of the elementary excitations in such dilute normal Fermi fluids at very low temperature that the present work is addressed.

We assume that the vapour has reached thermal equilibrium inside a magnetic trap and regard it as a dilute, polarized Fermi fluid. The $s$-wave collision between pairs of atoms are inhibited by the Pauli princile, so that to leading order only $p$-wave scattering and dipole magnetic interactions are important. However, these interaction effects are very small at the temperatures of present interest \cite{stoof}-\cite{maddalena}. The confined vapour may then be treated as a non-interacting Fermi gas, with resistance to shape deformation being provided by exchange in the regime of quantal degeneracy.

In Sect.~2 we start treating an inhomogeneous gas composed of (a large number of) non-interacting fermions from the general viewpoint of the equation of motion for the Wigner distribution function and of its projections yielding the conservation laws in the form of quantal hydrodynamic  equations \cite{bosoni_loro}. On neglecting temperature fluctuations, the kinetic stress tensor contains the essential physical input  for the study of both the equilibrium density profile and the dynamics of density fluctuations. We assume next that (i) the collision mechanisms that have brought the vapour to thermal equilibrium are still operative during its dynamics, and (ii) the inhomogeneity is weak on the lenght scale set by the inverse Fermi wavenumber. Under these approximations the kinetic stress tensor is simply related to the local Pauli pressure and to the velocity field. We recover from this approach the Thomas-Fermi form of the equilibrium density profile already found by M{\o}lmer \cite{molmer} for fermions in a boson-fermion mixture at zero temperature. We proceed from here in Sect.~3 to evaluate the eigenmodes of small-amplitude oscillations for the Fermi vapour at zero temperature in a spherically symmetric trap. In an Appendix we show how one may recover within the same approach the results of Stringari \cite{stringari_hydro_1996} for the collective excitations of a Bose condensate in the strong coupling regime and comparatively discuss the nature of the eigenmodes solutions for the Fermi gas and for the Bose condensate.

In Sect.~4 we extend the discussion of the small-amplitude vibrations of an inhomogeneous Fermi gas to the case of confinement within an axially symmetric harmonic trap and give explicit results for the low-lying eigenmodes. In Sect.~5 we consider instead the dynamics of a Fermi gas under spherical confinement in the case where collisions have a negligible effect on the deviations from the equilibrium. We make brief reference to zero-sound excitation of the homogeneous Fermi gas in such a collisionless regime and then focus on a variational treatment of a Landau-Boltzmann equation for the Wigner distribution function. Finally, Sect.~6  gives a brief summary of our main results and offers some concluding remarks.

\section{Small amplitude excitations in the collisional regime}

We describe the inhomogeneous spin-polarized fermion cloud by the one-particle 
density matrix 
\be
\rho(\vett{x},\vett{x}';t)=\langle \hat \Psi^{\dag}(\vett{x},t) \hat\Psi(\vett{x}',t)\rangle
\label{matrix}
\ee
where $\hat \Psi$ and $\hat \Psi^{\dag}$ are the fermionic field operators 
satisfying  the usual anticommutation relations, and we have suppressed the spin index, which is frozen because the magnetically-trapped fermions belong to a single hyperfine level. The equation of motion for the density matrix of non-interacting fermions in an external potential $V_{ext}(\vett {x},t)$ is  (see for istance Singwi and Tosi \cite{singwi}):
\be
\label{due}
\left[i \hbar \partial_t - \frac{\hbar^2}{2m}(\nabla^2_{\vett {x}}-\nabla^2_{\vett{x}'})+V_{ext}(\vett{x},t)-V_{ext}(\vett{x}',t)\right]\rho(\vett{x},\vett{x}';t)=0
\ee
We shall take
$V_{ext}(\vett{x},t)$  as the sum of a static harmonic  potential $V_0(\vett{x})$ and the perturbing potential $V_1(\vett{x},t)$.

The generalized hydrodynamic equations expressing the conservation laws are obtained by expanding Eqn.~\refeq{due} in powers of $\vett{r}=\vett{x-x'}$ around the diagonal $\vett{R}=(\vett{x}+\vett{x}')/2$. We introduce the Wigner distribution function which is related to $\rho(\vett{x},\vett{x}',t)$
 by:
\be
f_{\vett{p}}(\vett{R}, t)=\int d\vett{r} 
\rho(\vett{R}+\vett{r}/2, \vett{R}-\vett{r}/2;t) \exp(i\vett{p}\cdot \vett{r}/\hbar),
\ee
and its successive moments giving the density of particles $n(\vett{R},t)$,
\be
n(\vett{R},t)=\int \frac{d\vett{p}}{(2 \pi \hbar)^3}f_{\vett{p}}(\vett{R}, t),
\ee
the current density $\vett{j}(\vett{R},t)$,
\be
\vett{j}(\vett{R},t)=
\int \frac{d\vett{p}}{(2 \pi \hbar)^3}
\frac{\vett{p}}{m}f_{\vett{p}}(\vett{R}, t),
\ee
and the kinetic stress tensor $\Pi_{ij}(\vett{R},t)$,
\be
\Pi_{ij}(\vett{R},t)=
\int \frac{d\vett{p}}{(2 \pi \hbar)^3}
\frac{p_{i}p_{j}}{m}
f_{\vett{p}}(\vett{R}, t).
\ee
Then by the indicated expansion procedure we obtain the conservation laws in the form:
\be
\label{tre}
\partial_t n(\vett{R},t)=- \nabla \cdot j(\vett{R},t)
\ee
 and
\be
\label{quattro}
m \partial_t j_i(\vett{R},t), = - \nabla_j \Pi_{ij}(\vett{R},t)-n(\vett{R},t) \nabla_i V_{ext}(\vett{R},t)\;. 
\ee

We consider below only isothermal fluctuations and therefore omit to write down the energy conservation equation.

\subsection{\it Approximations on the kinetic stress tensor}
The kinetic stress tensor of an homogeneous fluid is related to the pressure $P$ and to the velocity field $\vett{v}$ by
\be
\Pi_{ij}= P\delta_{ij} + n m v_{i}v_{j}
\label{8}
\ee
Fore the ideal spin-polarized Fermi gas at zero temperature we have
\be
P=2E/3V = 3n\epsilon_{f}/5
\ee
where $E$ is the gound state energy and $\epsilon_{f}= \hbar^2 (6\pi^2 n)^{2/3}/2m$ is the Fermi energy.
On the other hand, the second term on the RHS of Eqn.~\refeq{8} can be dropped in the linear regime.

These results can be used to obtain the kinetic stress tensor of the inhomogeneous Fermi fluid if the inhomogeneity is sufficiently weak to allow a local-density approximation, {\it i.e.}
\be
\label{undici}
\Pi_{ij}(R,t)=\delta_{ij} P(R,t) = \delta_{ij} \frac{2}{5} A [n(R,t)]^{5/3} 
\ee
where $A= \hbar^2 (6 \pi^2)^{2/3}/2m$. More explicitly, we are assuming that the lenght scale for the variation of the density profile in space is large relative to the inverse Fermi wavenumber $k_{f}^{-1}$.

We remark that insertion of Eqn.~\refeq{undici} into Eqns.~\refeq{tre} and~\refeq{quattro} and reduction to the case of a static external potential yields the equation for the equilibrium density profile $n_{0}(\vett{R})$ in the form
\be
\frac{2}{5} A \nabla n_0^{5/3}+n_0 \nabla V_0=0.
\label{12}
\ee
The solution of Eqn.~\refeq{12} is the Thomas-Fermi profile  for the confined Fermi vapour:
\be
\label{tredici}
n_0(R)= 
\theta(\epsilon_{f} -V_0(\vett{R}))
\left(\frac{\epsilon_f-V_0(R)}{A}\right)^{3/2}
\ee
(see for instance M{\o}lmer \cite{molmer}). The profile vanishes continuosly at a radius $R_f$, say.

\subsection{\it Equation of motion for small amplitude fluctuations}
With regard to the dynamics of the fermionic cloud, the approximation \refeq{12} is most suitable for treating a ``collisional'' regime in which the relaxation of the fluctuations is rapid on the time scale set by the frequency of the driving potential. We shall have to return to this point in Sect.~5 below.
Here we proceed to study the dynamics of the fermion cloud for small distorsions of its density profile around the equilibrium Thomas-Fermi profile given in Eqn.~\refeq{tredici}.

If we denote by $n_{1}(\vett{R}, t)$ such a small-amplitude density fluctuation, Eqns.~\refeq{tre}-\refeq{quattro} and \refeq{undici} can be linearized and combined to yield the equation of motion
\be
\label{16}
\partial_t^2 n_1(\vett{R},t)=\frac{1}{m}\nabla_{\vett{R}} \cdot \left\{n_1(\vett{R})\nabla_{\vett{R}} V_0(\vett{R}) +
\frac{2}{3}A\nabla_{\vett{R}} \left[n_{0}^{2/3}(\vett{R})
n_1(\vett{R})\right]\right\}\;.
\ee
Equation \refeq{16} holds at resonance, since we have dropped the external perturbation drive.
In the homogeneous limit ($V_0(\vett{R}) \rightarrow 0$) it describes
the  sound propagation at a speed $c^2=2 \epsilon_f/3m$ -- a well-known result for first sound  in the ideal Fermi gas \cite{nozieres}. In the so-called collisional regime the local compression and expansion of the homogeneous fluid which are induced by the sound wave are reflected in spherically symmetric oscillations of the diameter of the Fermi sphere.

\section{Small oscillations in a spherical trap}
We determine in this section the solution of Eqn.~\refeq{16} for the case of a spherically symmetric harmonic confinement, {\it i.e.} $V_0(\vett{R})=m\omega_{f}^{2} R^{2}/2$.
We Fourier-transform Eqn.~\refeq{16} with respect to time and rescale the lengths with the harmonic oscillator length $a_{ho}=(\hbar/m\omega_f)^{1/2}$, by setting $\vett{R}=\vett{x} a_{ho}$.
We get:
\be
-3\omega^2 n_1(\vett{x},\omega)=\omega_f^2\nabla_{\vett{x}}\cdot
\left[n_1(\vett{x},\omega)\nabla_{\vett{x}}(x^2/2)+(X^2-x^2)\nabla_{\vett{x}}n_1(\vett{x},\omega)\right]
\label{eqmodinormali}
\ee
where $X=\sqrt{2\epsilon_F /m\omega_f^2}/a_{ho}$ is the scaled radius of the fermionic cloud. We search for spherical solutions of Eqn.~\refeq{eqmodinormali} having the form:
\be
n_1(\vett{x},\omega)= \theta(X-x)x^{l} F(x/X) Y_l^m(\theta,\phi)
\label{automodi}
\ee
with $Y_l^m(\theta,\phi)$ being the spherical harmonics.
Clearly, we have imposed on the eigenmodes solutions the condition that they 
be non-vanishing only inside the equilibrium cloud radius $X$.
In addition, we shall impose that the solutions vanish continuously at the 
cloud boundary $x=X$, as a consequence of Fermi statistics giving a high cost 
in kinetic energy to rapid variations of the density profile in space. This boundary condition on the eigenmodes of vibration for a trapped Fermi gas is further discussed in the Appendix and it is contrasted there with that imposed {\it e. g.} by Stringari \cite{stringari_hydro_1996} on those of a trapped (interacting) Bose gas, where considerations of kinetic energy only play a secondary role.

After setting $y=x/X$ and substituting  expression (\ref{automodi}) into Eqn.~(\ref{eqmodinormali}) we find the differential equation satisfied by $F(y)$:
\bea
(1-y^2)\frac{d^2}{dy^2}F(y) + 
\left(\frac{2(l +1)-(2l +3)y^2}{y}\right)\frac{d}{dy}F(y)
 +\nonumber \\
+
(3+3(\omega/\omega_{f})^{2}-l)F(y)=0.
\label{radialeq}
\eea
with the boundary condition $F(1)=0$.
We adopt the Fuchs method \cite{fuchs} for solving in a series form  ordinary differential equations around regular singular points. Using the above-mentioned condition of continuity at the boundary, and the fact that $F(y)$ is an even function of its argument, the solutions of Eqn.~\refeq{radialeq} are
\be
F(y)=(1-y^2)^{1/2}
\sum_{k=0}^{\infty} \alpha_{k} (1-y^2)^{k}
\label{20}
\ee
and the coefficients $\alpha_{k}$ satisfy the recurrence equation:
\be
\alpha_{k+1}=
\frac{(2k+1)(2k+2l+3) -(3+3(\omega/\omega_{f})^2 -l)}
{2(2k+3)(k+1)}\;\alpha_{k}
\label{ricorrenza}
\ee
Finally, the dispersion relation for the normal modes is obtained by imposing that the solutions \refeq{20}-\refeq{ricorrenza} reduce to polynomials, {\it i.e.}
$\alpha_{n+1}=0$ for an integer $n$ representing the number of internal nodes of the density fluctuation profile. This yields
\be
\omega(n,l)=\omega_{f}
\left(l+\frac{4}{3} n(n+l+2)\right)^{1/2}
\label{rel.disp.}
\ee
The dispersion relation \refeq{rel.disp.} correctly displays the sloshing mode solution 
($n=0$, $l=1$, $m=0$) at a frequency $\omega=\omega_{f}$, in agreement with 
the generalized Kohn theorem \cite{kohn}.

It is also interesting to notice that the eigenfrequencies surface modes at 
$n=0$ are the same as those found by Stringari \cite{stringari_hydro_1996} for
a Bose condensate in a spherical trap and that the $l=0$ modes resemble 
those found for a Fermi superfluid by Baranov and Petrov \cite{baranov}.

\section{Extension to axially-symmetric magnetic confinement}
We evaluate in this section the low-lying solutions of Eqn.~\refeq{16} in the case of axially symmetric harmonic confinement, which is more directly relevant in regard to experiment. The confining potential  is chosen as 
\be
V_0(\rho, z)=\frac{1}{2}m\omega_{f}^{2}(\rho^{2}+\lambda^{2}z^{2})
\ee
where $\rho=\sqrt{x^{2}+y^{2}}$ is the radial coordinate and $\lambda$ is the anisotropy parameter.

In the axially symmetric case only the  $L_z$ component of the angular momentum along the $z$ direction is conserved, whereas $l$ is no longer  a good quantum number. However, for the low-lying modes we can still start from a spherical base in  Eqn.~\refeq{20}, and suitably modify it so as  to account for the cylindrical symmetry of the problem. As we shall see through an example below, the consequence of the reduced symmetry is 
a coupling between some of the modes which are characterized by different values of $l$ in the spherical base.

We give here the explicit results for the eigenfrequencies and the density-fluctuation profiles in some istances of  dipolar and quadrupolar modes In the expressions reported below, all lengths are still scaled in units of $a_{ho}=(\hbar/m \omega_{f})^{1/2}$.

\subsection{\it Dipole sloshing mode in the $z$ direction.}
Starting from the spherical-base  solution corresponding to ($n=0$, $l=1$, $m=0$), we construct the density profile
\be
n_{1}^{(0,1,0)}(\rho, z)=(1-\rho^2 - \lambda^2 z^2)^{1/2}\rho
\ee
The corresponding frequency is 
$\omega^{2}=\lambda^2 \omega_{f}^2$
in agreement with the generalized Kohn theorem \cite{kohn}.

\subsection{\it Quadrupole surface modes. }

The anisotropy introduces a splitting between the quadrupolar modes corresponding to ($n=0,l=2,m=1$) and to ($n=0,l=2,m=2$). the density profiles and the eigenfrequencies are as follows:

\bea
n_{1}^{(0,2,1)}(\rho, z)= (1-\rho^2 - \lambda^2 z^2)^{1/2}\rho z\exp(i\phi)
\\
\omega=(1+\lambda^2)^{1/2} \omega_{f}
\eea 

\bea
n_{1}^{(0,2,2)}(\rho, z)= (1-\rho^2 - \lambda^2 z^2)^{1/2}\rho^{2}\exp(2i\phi)
\\
\omega=2^{1/2} \omega_{f}
\eea 
In these equations $\phi$ is the angular momentum variable in the transverse plain. The two eigenfrequencies become degenerate in the spherical limit.

\subsection{\it Monopole and quadrupole coupled modes.}
The anisotropy introduces a coupling between the modes described by the quantum numbers ($n=0,l=2,m=0$) and ($n=1,l=0,m=0$). The associated density fluctuations are
\be 
n_{1}^{(0,2,0)}=(1-\rho^2 - \lambda^2 z^2)^{1/2}(2z^2 -\rho^2)
\ee
 and
\be
n_{1}^{(1,0,0)}=(1-\rho^2 - \lambda^2 z^2)^{1/2}(a - \rho^2 - z^2)
\ee
where $a$ has to be determined by a normalization condition. For the coupled eigenmode equations we find the eigenfrequencies as:
\be
\omega^{2}= \omega_{f}^{2}
\left(
\frac{4 \lambda^{2} +5 \pm \sqrt{16 \lambda^{4}-32\lambda^{2} +25}}
{3}\right).
\ee

The above examples should be sufficiently illustrative of the procedure to be followed in determining the small-amplitude oscillations in the case of axially anisotropy.

\section{Collisionless regime}
\label{subsec2}

We return to the case of spherical confinement for the purpose of discussing the role of collisions in the dynamics of the fermion cloud.
We have already remarked in Sect.~2 that in the homogeneous limit Eqn.~(\ref{16}), which was derived with the help of Eqn.~(\ref{undici}) for the kinetic stress tensor, describes first sound in a collisional regime where the local Fermi sphere executes a ``breathing'' oscillation. In the so-called collisionless (or zero-sound) regime, on the other hand, the relaxation time of density fluctuations in the homogeneous Fermi fluid is long compared with the period of the eigenmode and the local Fermi sphere undergoes anisotropic oscillations (see for istance Pines and Nozi{\`e}res \cite{nozieres}). In the limit of vanishing coupling the speed of the zero sound tends to the Fermi velocity $v_{F}$ (compared with the value $3^{-1/2}v_{F}$ for the velocity of the first sound) and the deformation of the Fermi surface reduces to a small bump in the direction of the propagation vector. The collective zero-sound mode involves in this limit only a small number of quasiparticles and essentially propagates at their velocity.

For a discussion of the dynamics of a magnetically confined Fermi cloud  in a regime where collisions have little influence, we resort to an equation for the Wigner distribution function having the form of a Boltzmann equation with the collision integral set to zero, that is
\begin{equation}
\label{bz}
\partial_t f_{\vett p}(\vett R,t)+\frac{\vett p}{m} \cdot \nabla_{\vett R} f_{\vett p}(\vett R,t)-\nabla_{\vett R} V_0(\vett R) \cdot\nabla_{\vett p} f_{\vett p}(\vett R,t)=0 
\end{equation}
This equation is in fact analogous to the Landau transport equation for quasiparticles in a homogeneous Fermi fluid in the collisionless regime and is expected to be valid down to very low temperature provided that the cloud contains a large number of fermions.

In searching for solutions of equation (\ref{bz}), we focus on the
shape-deformation modes which can be monitored by measuring the mean
square radius of the cloud as a function of time. Within this class of
modes, we adopt a variational {\it Ansatz} as already proposed by
Bijlsma and Stoof \cite{michiel1,michiel2} for a Bose cloud. This amounts to choosing  $f_{\vett p}(\vett R,t)$ in the form of the equilibrium solution $\Phi_{eq}(\vett p,\vett R)$ after rescaling both the space and momentum variable through time-dependent factors in each geometrical direction. More precisely, for ($i=x,y,z$)  we introduce variational parameters $\alpha_i(t)$ describing the deviation of the mean square sizes of the cloud from their equilibrium values,
\be
\alpha_i(t)=[\langle R_i(t)^2 \rangle/\langle R_i^2\rangle_{eq}]^{1/2}
\ee
The form of $f_{{\vett p}}({\vett R},t)$ is taken as
\be
\label{**}
f_{{\vett p}}({\vett R},t)
=( c_x c_y c_z) \Phi_{eq}\left(\sqrt{c_{i}}\alpha_i(p_{i}-mR_{i}\dot{\alpha_{i}}/\alpha_{i}) ;\; 
\sqrt{c_i}R_i/\alpha_{i} \right)\;.
\ee
the time-independent coefficients 
 $c_i$ being introduced in Eqn.~\refeq{**} in order to ensure  that the initial momentum distribution is isotropic.
The equilibrium form of the Wigner function for the degenerate Fermi gas subject to spherical confinement  can in turn be chosen as a generalized Fermi sphere:
\be
\Phi_{eq}(\vett p,\vett R)=\theta\left(
\frac{p^2}{2m} + \frac{1}{2}m \omega_{f}^{2}R^2 -\mu_{f} \right)
\label{phieq}
\ee
$\mu _{f}$ being the chemical potential.
Of course, Eqn.~(\ref{phieq}) yields the same equilibrium densitty as that reported earlier in eq. (\ref{tredici}). We emphasize that in the present treatment we are assuming that, even though collisions have negligible influence on the dynamics of the cloud, they must have been active in the past in order to bring the cloud to the state of thermal equilibrium.

The equations of motion for the $\alpha_{i}(t)$ in each spatial
direction are obtained 
by taking the moments of equation (\ref{bz})with respect to $R_ip_i$ ~\cite{michiel1,michiel2}:
\be
\ddot{\alpha_{i}} +\omega_{f}^{2} \alpha_{i} =\frac{\omega_{f}^{2}}{\alpha_{i}^{3}}.
\label{eq_alpha}
\ee
It is worth noticing that for a non-interacting gas Eqn.~\refeq{eq_alpha} follows directly from the dynamical scaling {\it ansatz}~\refeq{**} and from the virial theorem, and does not require a specific assumption on the shape of the variational function.

The solutions of the non-linear equation of motion (\ref{eq_alpha}), under the initial condition $\alpha_{i}(0)=1$, is
\be
\alpha_{i}(t)=\frac{1}{\sqrt{2}~ \omega_{f}}
[c+(c^2-4\omega_{f}^{4})^{1/2}
\sin ( \pm 2 \omega_{f}t-\phi)]^{1/2}
\ee
with $\sin(\phi)=[(c-2 \omega_{f}^{2})/(c+2 \omega_{f}^{2})]^{1/2}$ and the constant $c$ determined by the initial velocity according to:
\be
\dot \alpha (0)=\pm (c-2 \omega_{f}^2)^{1/2}
\ee

A more direct contact can be made with  the results reported in the preceding sections for the gas in the collisional regime by linearizing Eqn.~\refeq{eq_alpha}. In such small-oscillation regime, Eqn.~\refeq{eq_alpha} becomes an equation for the linearized mean square radius $\alpha^{(1)}_i(t)$:
\be
\ddot \alpha^{(1)}_{i}+4\omega_{f}^{2}\alpha^{(1)}_{i}=0\;.
\ee
Evidently one obtains an oscillation of the cloud at 
 frequency $\omega=2\omega_{f}$, corresponding to the frequency of a single-particle excitation in the cloud.

As a final remark we point out that while Eqn.~\refeq{bz} satisfies the generalized Kohn theorem, a more sophisticated variational {\it ansatz} would be needed to extract from it the sloshing mode \cite{michiel2}.

\newpage
\section{Summary and concluding remarks}

In summary, we have presented an investigation of the collective excitations of a spin-polarized Fermi gas confined in an external harmonic trap. We have obtained analytic results for the small-amplitude eigenmodes in a dynamical regime corresponding to first sound in the bulk and proposed a Landau-Boltzmann approach to treat the dynamics of the confined gas in a dynamical regime corresponding to zero sound in the bulk. In spite of the assumed negligible role of interactions between the particles of the gas in contributing to a mean field  sustaining a collective oscillation, we have seen that the Pauli pressure alone can determine a rather rich dynamical behaviour.

Trapped mixtures of bosonic and fermionic species are expected to become accessible to experiment in near furure. The investigation of collective excitations in such mixtures seems a natural direction of development for the present study. The density profiles of the mixtures have been determined theoretically both in the ground state at zero temperature \cite{molmer} and in the equilibrium state at finite temperature \cite{cataliotti}. From these studies one can easily envisage situations in which the dynamics of a boson-fermion mixture should be amenable to analytical approaches. Theoretical progress in this area will be reported elsewhere.

\subsection*{Acknoledgments}
This work is supported by the Istituto Nazionale di Fisica della Materia throught the Advanced Research Project on BEC. One of us (I.M.) wishes to thank Professor M. Fontana and the Istituto Nazionale di Fisica della Materia for a short-term grant which has permitted her partecipation at this work.

\newpage

\appendix

\section*{Appendix A. Collective excitations of Fermi  versus  Bose-condensed clouds}

 In this Appendix we first show how the dynamics of  a dilute Bose
 condensed cloud at zero temperature can be studied within the
 Wigner-distribution  formalism that we have used for the
 spin-polarized Fermi gas (see also \cite{bosoni_loro}). We thereby
 derive the analogue of the equation of motion (\ref{16}) for density
 fluctuations in the Bose-condensed cloud and point out that for
 harmonic confinement it coincides with the one obtained by Stringari
 \cite{stringari_hydro_1996} in a hydrodynamic approach to the strong
 coupling limit. We then comparatively discuss the nature of the
 solutions of the two equations of motion for the Fermi gas and for
 the Bose-condensed cloud.

\subsection*{A1. Equation of motion for density fluctuations in a Bose-condensed cloud}
In the case of a Bose-condensed system at zero temperature, and
neglecting depletion due to the interactions, the one-body density matrix
introduced in Eqn.~(\ref{matrix}) reduces to the simple product of two condensate
wave functions,
\be
\langle \Psi^{\dagger}(\vett x,t) \Psi(\vett x', t)\rangle=
\Phi^{*}(\vett x, t) \Phi(\vett x', t).
\ee
	   						
The generalized hydrodynamic equations (\ref{tre}) and (\ref{quattro}) are thereby closed
and the kinetic stress tensor is explicitly given by
\be
\Pi_{ij}(\vett R,t) =
\frac{\hbar^{2}}{m}
\left[
\frac{\partial}{\partial r_{i}}
\frac{\partial}{\partial r_{j}}
\Phi^{*}(\vett R-\frac{\vett r}{2}, t) \Phi(\vett R+\frac{\vett r}{2}, t)
\right]_{\vett r=0}.
\ee

However, for a Bose condensate with repulsive interactions in the strong
coupling limit the kinetic stress tensor becomes in fact negligible
compared with the confinement and interaction energy terms. The linearized
equation of motion for density fluctuations becomes
\cite{march&tosi,nuovocim}
\begin{eqnarray}
\partial_t^2 n_1({\vett R},t)=\frac{1}{m}\int d{\vett x}'\int {d\vett
x}''[n_0({\vett R}) \nabla^2_{\vett R}  \delta ({\vett R}-{\vett x}')
\nonumber \\
+ \nabla_{\vett R}n_0({\vett R}) \cdot \nabla_{\vett R}\delta({\vett R}-{\vett x}')]\nu({\vett x}',{\vett x}'')n_1({\vett x}'',t)
\label{deibose}
\end{eqnarray}

where $\nu({\vett x}',{\vett x}'')$  is the interparticle potential.

	Assuming contact interactions, i.e. $\nu({\vett x}',{\vett
x}'')=g\delta({\vett x}'-{\vett x}'')$  with  $g=4\pi\hbar^2a/m$ where $a$ is the s-wave
scattering length, Eqn.~(\ref{deibose}) immediately yields
\be
\partial_t^2n_1({\vett R},t)=\frac{g}{m}\nabla_{\vett R}\cdot
[n_0({\vett R})\nabla_{\vett R}n_1({\vett R},t)]
\label{string}
\ee

This yields the equation of motion derived by Stringari \cite{stringari_hydro_1996} when the
Thomas-Fermi solution is used for the equilibrium density profile
$n_0({\vett R})$.
	Before proceeding we wish to emphasize that the nature of the
Thomas-Fermi approximation is very different in the two cases that we are
considering. In the spin-polarized Fermi gas the interactions are
negligible, so that the kinetic term provides the only energy scale. On the
other hand, in a mesoscopic Bose condensate the interactions are dominant
even at zero temperature. Nevertheless, the form of Eqn.~(\ref{string}) is
essentially similar to that of Eqn.~(\ref{16}).

\subsection*{A2. Comparative discussion of the solutions of Eqn.~(\ref{string}) and Eqn.~(\ref{eqmodinormali})}
	The solutions of Eqn.~(\ref{string}) for the Bose-condensed cloud vanish
outside the cloud radius $R_b$ and present a discontinuity at $R_b$ \cite{stringari_hydro_1996}
. Indeed, by
the Fuchs method used in Section 3 one may prove that it is not possible to
impose continuity at $R_b$. The discontinuity is physically acceptable in view
of the fact that the kinetic energy term has been set as negligible in
taking the strong-coupling limit.
	On the other hand, Eqn.~(\ref{16}) for the Fermion cloud admits, in
addition to the solution that we have presented in the main text, a
non-vanishing solution outside the cloud radius $R_f$. This solution
	is
\be
n_({\vett R},\omega)=C \theta(y-1)y^{-(3+(\omega/\omega_{f})^2)}.
\label{dueA}
\ee

Evidently, such a finite external solution going to zero only for $y
\longmapsto \infty$ as in
Eqn.~(\ref{dueA}) is inconsistent with the assumption of small-amplitude
oscillations of the cloud. Nevertheless, if for a moment one admits it and
asks that the constant $C$  be finite and determined by imposing continuity at
the boundary with the internal solution (as necessary in a system where the
kinetic energy is dominant), one finds a dispersion relation which is
different from that given in Eqn.~(\ref{rel.disp.}), that is
\be
3(1+\omega^2/\omega_{f}^2)=l+4n(n+l+1).
\label{treA}
\ee	   						

It is evident that the modes with $n=0$  and $l\leq 3$ are suppressed, and in particular
the sloshing mode which is expected from the generalized Kohn theorem is
absent. We have for this reason discarded this second solution and, by
imposing that the internal solution vanishes continuously at the boundary
(i.e. $C=0$ ), found the dispersion relation reported in Eqn.~(\ref{rel.disp.}) in the main
text. We also remark that each physically acceptable eigenfrequency of the
cloud, with increasing $n$  at fixed $l$ in Eqn.~(\ref{rel.disp.}), lies between two of the
unphysical frequencies given by Eqn.~(\ref{treA}).
	In conclusion, the different roles played by the kinetic energy in
the Bose condensate and in the Fermi gas lead to very different forms of
the density fluctuations in these two systems. In the former system the
amplitude of the fluctuations vanishes discontinuously at the boundary of
the cloud in the strong-coupling limit, whereas in the latter it vanishes
continuously at the boundary.

\bibliographystyle{prsty}

\end{document}